\pdfoutput=1
\documentclass[prl,twocolumn,showpacs,amsmath,amssymb,floatfix,superscriptaddress]{revtex4-1}

\usepackage{graphicx}
\usepackage{dcolumn}
\usepackage{bm}
\usepackage{hyperref}

\begin{document}

\title{Correlation-driven topological phase transition from in-plane magnetized quantum anomalous Hall to Mott insulating phase in monolayer transition metal trichlorides}

\author{Xian-Lei Sheng}
\affiliation{Department of Physics and Astronomy, University of Delaware, Newark, DE 19716-2570, USA}
\affiliation{Department of Applied Physics, Beihang University, Beijing 100191, China}
\author{Branislav K. Nikoli\' c}
\email{bnikolic@udel.edu}
\affiliation{Department of Physics and Astronomy, University of Delaware, Newark, DE 19716-2570, USA}

\begin{abstract}
Based on density functional theory (DFT) calculations, we predict that a monolayer of OsCl$_3$---a layered material whose interlayer coupling is weaker than in graphite---possesses a quantum anomalous Hall (QAH) insulating phase generated by the combination of honeycomb lattice of osmium atoms, their strong spin-orbit coupling (SOC) and ferromagnetic ground state with {\em in-plane} easy-axis. The band gap opened by SOC is \mbox{$E_g \simeq 67$ meV} (or \mbox{$\simeq 191$ meV} if the easy-axis can be tilted out of the plane by an external electric field), and the estimated Curie temperature of such {\em anisotropic planar rotator} ferromagnet is $T_\mathrm{C} \lesssim 350$ K. The Chern number $\mathcal{C}=-1$, generated by the manifold of Os $t_{2g}$ bands crossing the Fermi energy, signifies the presence of a single chiral edge state in nanoribbons of finite width, where we further show that edge states are spatially  narrower for zigzag than armchair edges and investigate edge-state transport in the presence of vacancies at Os sites. Since $5d$ electrons of Os exhibit {\em both} strong SOC and moderate correlation effects, we employ DFT+U calculations to show how increasing on-site Coulomb repulsion $U$: gradually reduces $E_g$ while maintaining $\mathcal{C} = -1$ for $0 < U < U_c$; leads to metallic phase with $E_g = 0$ at $U_c$; and opens the gap of topologically trivial Mott insulating phase with $\mathcal{C}=0$ for  $U > U_c$. 
\end{abstract}

\maketitle

\textit{Introduction.---}The quantum anomalous Hall (QAH) insulator is a recently discovered~\cite{Liu2016,Kou2015} topological electronic phase where strong spin-orbit coupling (SOC) and ferromagnetic ordering lead to band gap $E_g$ in the bulk of a two-dimensional (2D) electron system, as well as conducting (i.e., gapless) chiral edge states at its boundaries. Its  topologically nontrivial band structure is characterized by a nonzero Chern number $\mathcal C$ counting the number edge states whose energy-momentum dispersion threads the gap of finite-width wires, while their wave functions have finite spatial extent around the wire edges. Experimental confirmation of QAH insulators is based on the observation of quantized Hall conductance in the absence of any external magnetic field~\cite{Kou2015}. 

Unlike closely related quantum hall (QH) insulator, where chiral  edge states allow spin-unpolarized electron to propagate in only one direction, or quantum spin Hall (QSH) insulator, where helical edge states appear in pairs with different chirality and spin polarization~\cite{Kane2005} due to preserved time-reversal symmetry, the edge states of QAH insulator allow only one spin species to flow unidirectionally. Thus, the edge state transport in nanowires made of QAH insulator is robust against {\em both} magnetic and nonmagnetic disorder, which makes them superior to edge states of QSH insulator where electrons can be backscattered by disorder (such as magnetic impurities) breaking the time-reversal symmetry. The QAH insulator is also superior in potential applications to QH insulator since the latter requires strong external magnetic field and typically very low temperatures. 

\begin{figure}
	\includegraphics[scale=0.35,angle=0]{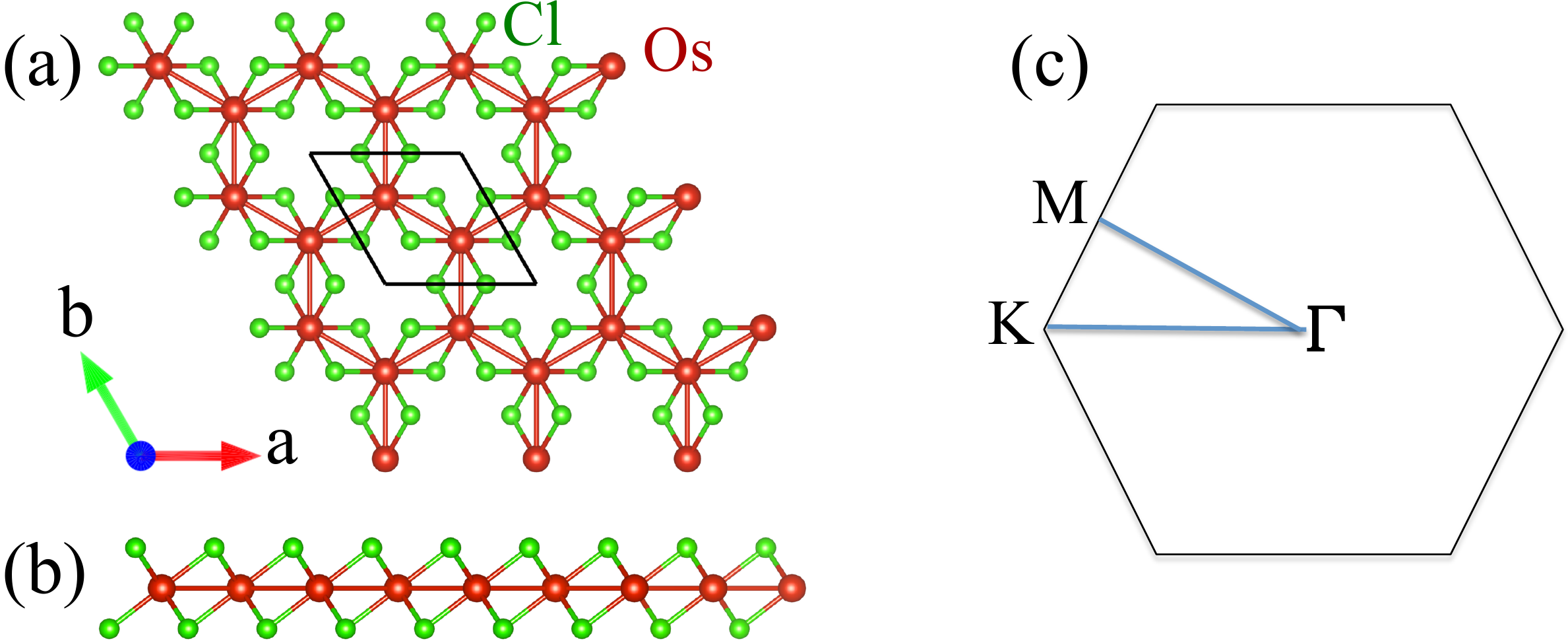}
	\caption{(Color online) (a) Top and (b) side view of the crystalline structure of a monolayer OsCl$_3$ where a sheet of transition metal Os atoms is sandwiched between two sheets of Cl atoms. The Os atoms form a honeycomb lattice, and each Os atom is located at an octahedral site between the Cl atom sheets. (c) First Brillouin zone of the monolayer OsCl$_3$ with high-symmetry points denoted.}
	\label{fig:fig1}
\end{figure}

However, QAH insulators have been observed thus far only at ultralow temperatures $\lesssim 100$ mK~\cite{Liu2016,Kou2015}, thereby igniting intense theoretical and computational searches~\cite{Dolui2015} for systems whose {\em both} $E_g$ and Curie temperature $T_\mathrm{C}$ are higher than the room temperature. Finding such materials, or heterostructures of different materials~\cite{Qiao2014,Zhang2015a}, would open new avenues for  nanelectronic and spintronic devices with  ultralow dissipation where edge states  act as ``chiral interconnects'' whose 
resistance is independent of the length of the wire~\cite{Liu2016}.

The seminal work on the Haldane model~\cite{Haldane1988}, where quantized Hall conductance is found for an electronic system defined on the honeycomb lattice with SOC in the absence of an external magnetic field, has inspired search for realistic materials exhibiting QSH or QAH insulating states where the honeycomb lattice structure is often the first ingredient~\cite{Wu2014,Xiao2011}. For example, graphene with  enhanced intrinsic SOC~\cite{Kane2005,Weeks2011,Gmitra2016} can be transformed into QSH insulator, and then converted into QAH insulator by adding exchange magnetic field (via impurities, doping or proximity effect) in order to suppress one of the two helical edge states of QSH insulator. Systems predicted to realize this concept include graphene decorated with heavy $5d$ transition metal adatoms~\cite{Zhang2012c}, or heterostructures like graphene/G-type-antiferromagnet~\cite{Qiao2014} and graphene/ferromagnetic-insulator~\cite{Zhang2015a,Wang2015c}. The proximity SOC in graphene can be further enhanced by inserting a monolayer of transition metal dichalcogenides (TMD) into such heterostructures~\cite{Gmitra2016}. Similarly, honeycomb lattice of silicene, germanene and stanene, which already possess strong intrinsic SOC, could be converted into QAH insulator by introducing exchange interaction via magnetic  adatoms~\cite{Zhang2013b} or surface functionalization~\cite{Wu2014}. Finally, honeycomb lattice enforcing trigonal symmetry of the crystalline field can introduce additional level splitting of the $d$-orbitals, which combined with SOC makes possible opening of topological gaps in transition metal oxide heterostructures~\cite{Xiao2011,Doennig2016}. However, most of the heterostructures have $E_g$ below room temperature, thereby requiring strain engineering or external pressure to increase $E_g$~\cite{Qiao2014}. Also, despite numerous predictions~\cite{Weeks2011,Zhang2012c,Zhang2013b} for topological phases in graphene decorated by heavy adatoms none has been realized thus far (in part, due to difficulties in keeping adatoms sufficiently apart from each other~\cite{Cresti2014,Chang2014}). 

\begin{figure}
	\includegraphics[scale=0.36,angle=0]{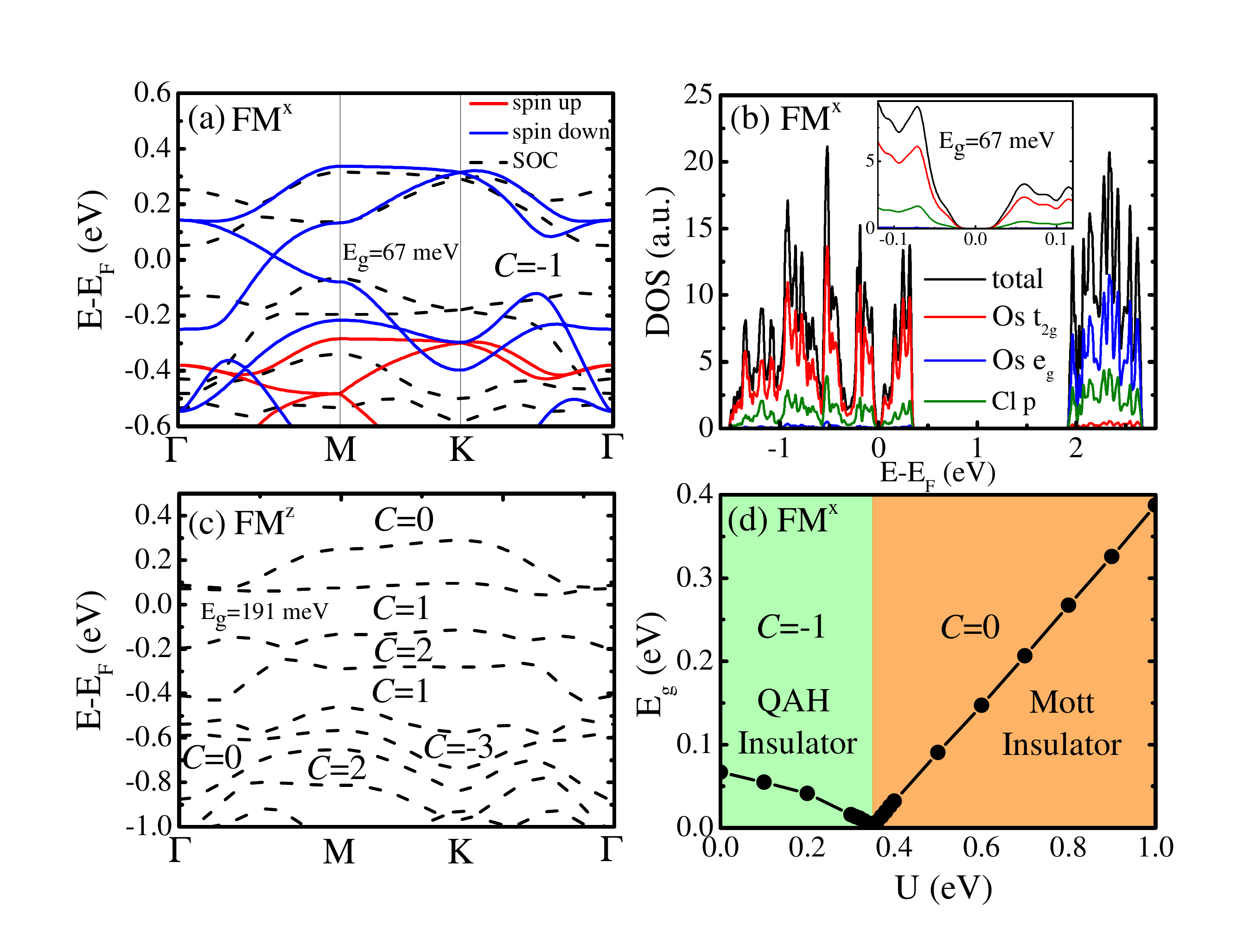}
	\caption{(Color online) Electronic band structure of monolayer OsCl$_3$ with Os spins in the: (a) in-plane FM$^x$ configuration shown in Fig.~\ref{fig:fig4}(a); and (c) out-of-plane FM$^z$ configuration (see also Table~\ref{tab:magnet}). The band structure is computed with GGA (solid lines) or GGA+SOC (dashed lines). (b) Density of states corresponding to panel (a). (d) The phase diagram of OsCl$_3$ with Os spins in the FM$^x$ configuration calculated by GGA+SOC+U.}
	\label{fig:fig2}
\end{figure}
\begin{figure}
	\includegraphics[scale=0.34,angle=0]{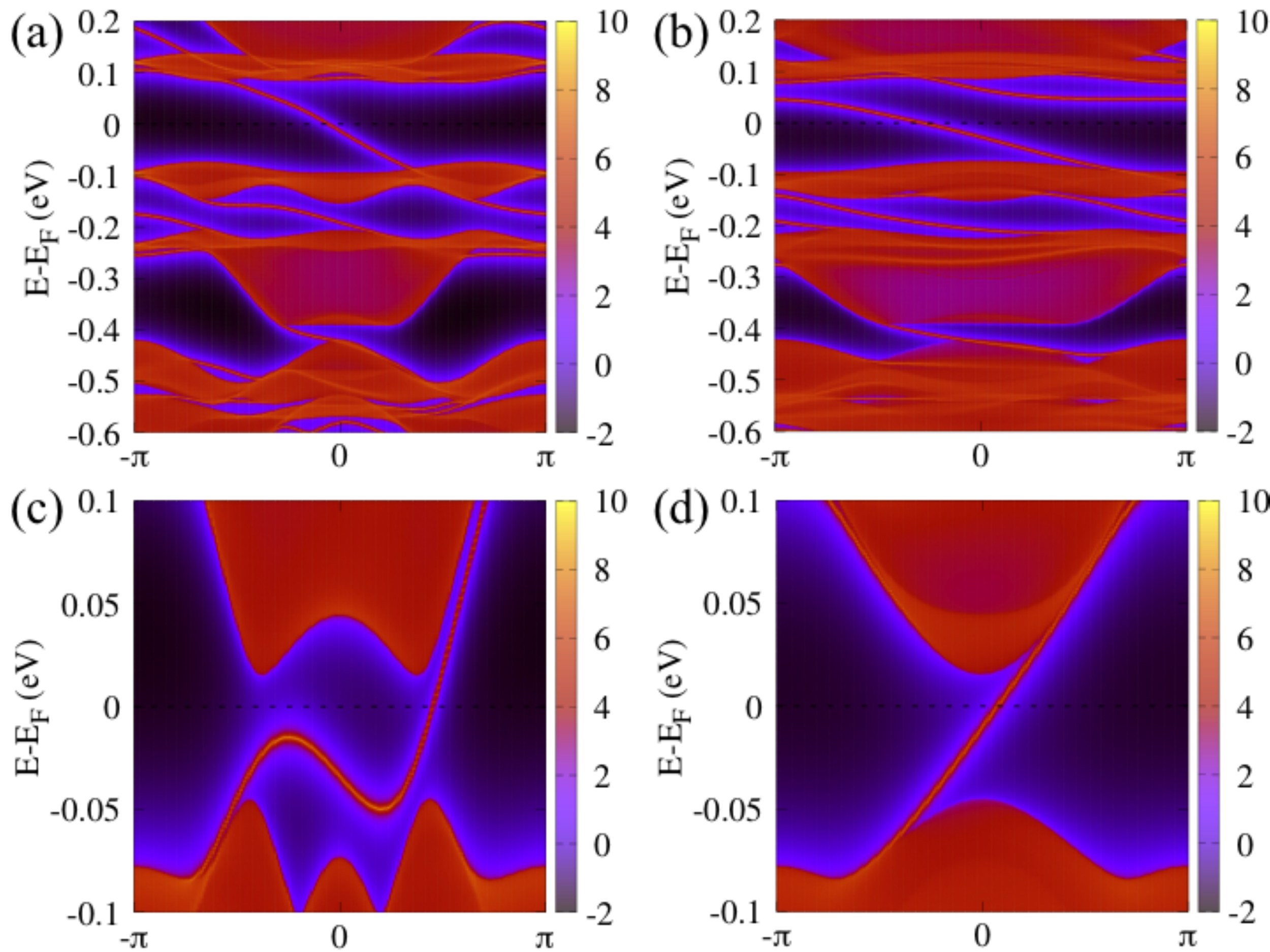}
	\caption{(Color online) Energy and $k_x$ dependence of the local DOS on the edge of semi-infinite sheet of OsCl$_3$ which is terminated by either (a),(c) zigzag or (b),(d) armchair arrangement of Os atoms along the edge. The Os spins are in FM$^z$ configuration in (a) and (b) or FM$^x$ configuration [illustrated in Fig.~\ref{fig:fig4}(a)] in (c) and (d). Warmer colors represent higher local DOS, where solid red regions indicate bulk energy bands and the blue regions indicate bulk energy gap.}
	\label{fig:fig3}
\end{figure}

\begin{table}[b]
	\caption{Interlayer binding energy  $E_b$ (in meV/\AA$^2$) of selected transition metal trichlorides MCl$_3$. For comparison, $E_b$  of typical layered materials like graphite or Bi$_2$Se$_3$ is also included. Note that computed value for graphite is comparable to experimental estimate \mbox{$E_b=23.3 \pm 1.9$} meV/\AA$^2$~\cite{Zacharia2004}.}\label{tab:tab1}
	\begin{tabular}{ccccccc}
		\hline 
		& OsCl$_3$ &  VCl$_3$  & FeCl$_3$  & RuCl$_3$ & graphite & Bi$_2$Se$_3$ \\
		\hline \hline
		$E_b$  & 14.4 &  17.3 & 18.1 & 19.3 & 26.4 & 23.9 \\ 
		\hline 
	\end{tabular}
\end{table}

Thus, recent efforts have also explored possibility of {\em intrinsic} QAH insulator realized in a single material~\cite{Dolui2015} with inverted band structure and ferromagnetic insulating behavior, but the latter is quite rare at room temperature~\cite{Liu2016}. In this Letter, we predict that a monolayer of OsCl$_3$ possesses an intrinsic QAH insulating phase characterized by an energy gap $E_g \simeq 67$ meV, opened only when SOC is turned on in our density functional theory (DFT) calculations, and Chern number $\mathcal{C} = -1$. The layered nature of several transition metal trichlorides MCl$_3$---where M = Ti, V, Cr, Fe, Mo, Ru, Rh, Ir---has been explored long before~\cite{Bengel1995} the present focus on the layered materials like graphene, TMDs and Bi-based three-dimensional topological insulators. The single layer of MCl$_3$ consists of Cl-M-Cl sandwich where a sheet of M atoms is sandwiched between two sheets of Cl atoms, with edge-sharing OsCl$_6$ octahedra forming a honeycomb lattice, as illustrated in Fig.~\ref{fig:fig1}. The weakness of van der Waals forces holding layers of MCl$_3$ together is illustrated in Table~\ref{tab:tab1} where we compare their interlayer binding energy, defined as $E_b=(E_\mathrm{monolayer}-E_\mathrm{crystal}/n)(|\vec{a}\times\vec{b}|)^{-1}$, with that of popular layered materials like graphite and  Bi$_2$Se$_3$. Here $E_\mathrm{monolayer}$ is the total energy of a monolayer unit cell, $E_\mathrm{crystal}$ is the total energy of a three-dimensional crystal unit cell, $n$ is the number of layers in a three-dimensional crystal unit cell and $|\vec{a}\times\vec{b}|$ is the area of the unit cell. 

\begin{figure}
	\includegraphics[scale=0.18,angle=0]{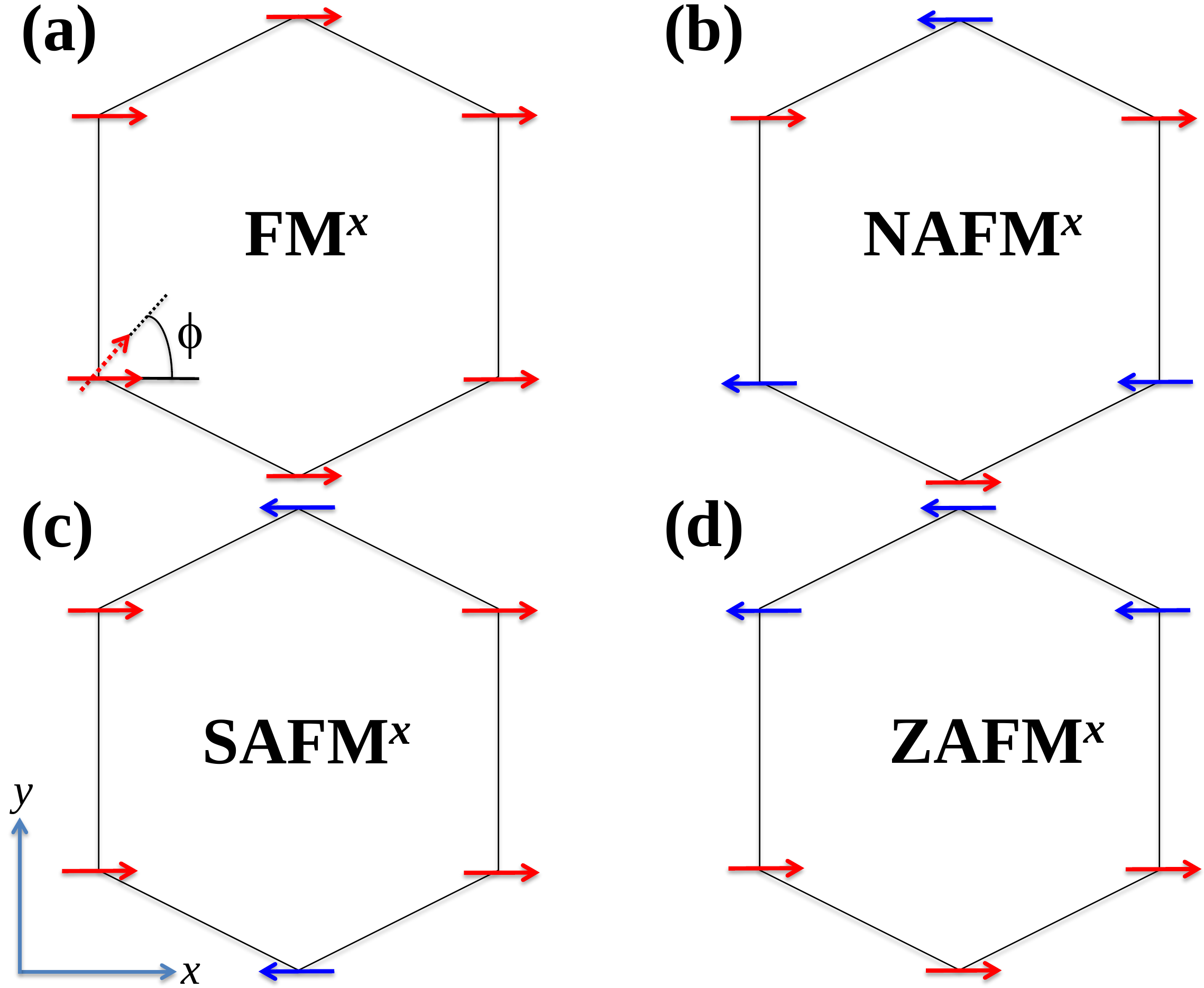} \hspace{0.1in} \includegraphics[scale=0.15,angle=0]{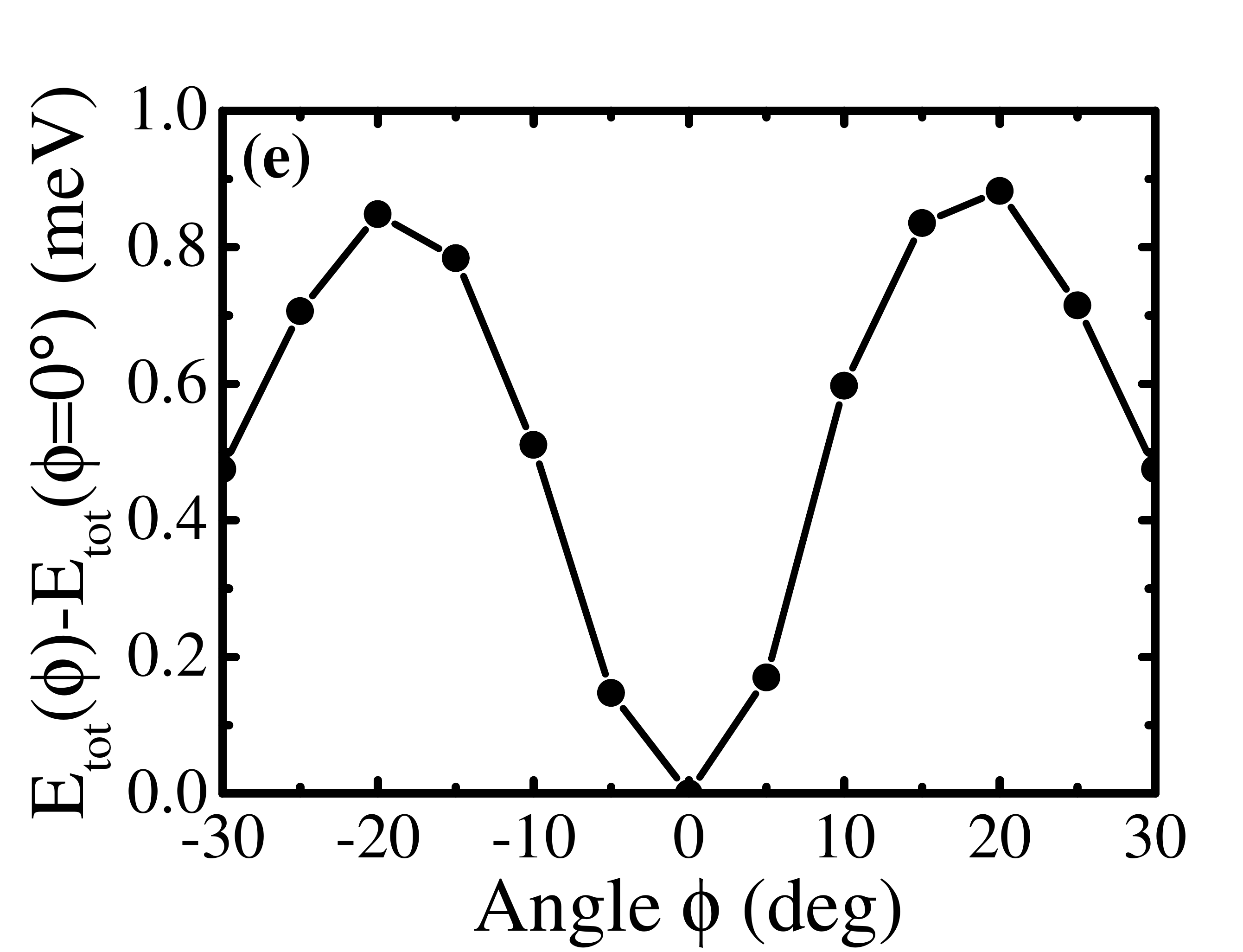}
	\caption{(Color online) Four possible magnetic configurations of Os spins: (a) ferromagnet (FM); (b)  N\'{e}el antiferromagnet (NAFM); (c) stripy AFM (SAFM); and 
		(d) zigzag AFM (ZAFM). (e) Change of total energy $E_\mathrm{tot}$ as spins in FM$^x$ configuration in (a) rotate within the $xy$-plane by an angle $\phi$.}
	\label{fig:fig4}
\end{figure}

\begin{figure}
	\includegraphics[scale=0.36,angle=0]{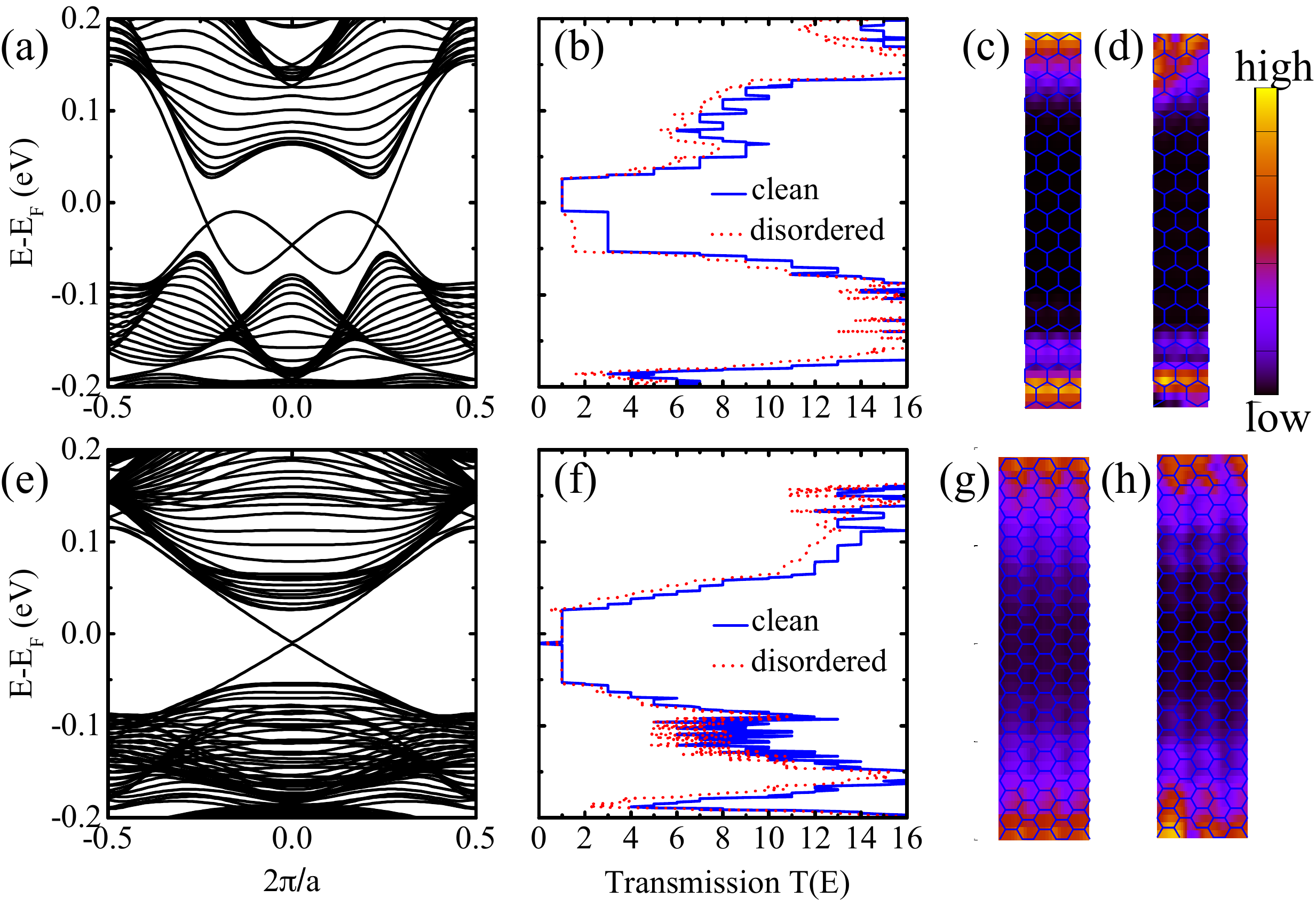}
	\caption{(Color online) Electronic subband structure of OsCl$_3$ nanoribbons for: (a) zigzag edges and width $W \simeq 10$ nm; (e) armchair edges and $W \simeq 12$ nm. The zero-bias transmission function $T(E)$ for zigzag and armchair nanoribbons, which are clean [as in panels (a) and (b)] or disordered by one vacancy at Os site on each edge are shown in panels (b) and (f), respectively. The local DOS for zigzag (armchair) nanoribbon with and without vacancy is shown in panels (c) and (d) [(g) and (h)], respectively.}
	\label{fig:fig5}
\end{figure}

We note that OsCl$_3$ is {\em virtually unexplored} among transition metal trichlorides where, e.g., monolayer RuCl$_3$ as SO-coupled Mott insulator~\cite{Rau2016} has recently been under intense scrutiny as a possible realization of the Kitaev spin-liquid phase~\cite{Jackeli2009,Price2012}. The topologically nontrivial band structure of OsCl$_3$ is brought about by the combination of honeycomb lattice formed by Os ions, strong SOC~\cite{Zhang2012c} of heavy transition metal osmium atoms and ferromagnetic ordering of their magnetic moments. Our predictions are based on first-principles calculations of electronic structure of an infinite sheet of  OsCl$_3$ in Fig.~\ref{fig:fig2}, as well as on tight-binding  calculations for semi-infinite sheets  where  Fig.~\ref{fig:fig3} confirm the presence of one chiral state per edge associated with $\mathcal{C}=-1$. While theoretical and computational searches~\cite{Liu2016,Dolui2015} for QAH insulators have largely been focused on finding materials or heterostructures with nonzero $\mathcal{C}$ of their bulk band structure,  we show that understanding of variety of possible spin orderings on the honeycomb lattice~\cite{Price2012} (see Fig.~\ref{fig:fig4}) or spatial and transport properties of chiral edge states (see Fig.~\ref{fig:fig5}) deserves their own attention.

In addition, the oxides of Ir~\cite{Wan2011,Witczak-Krempa2014,Rau2016} and Os~\cite{Wan2012} have recently emerged as a playground to study the interplay between strong SOC due to high atomic number  and moderate Coulomb interaction between electrons, which is weaker than in $3d$ transition metal oxides due to larger extent of $5d$ orbitals. Motivated by the fact that comparable energy scales associated with SOC and Coulomb interaction were found to lead to a variety of competing emergent quantum phases~\cite{Witczak-Krempa2014,Rau2016,Wan2012} in Ir and Os oxides, here we conduct analogous  analysis by using DFT combined with Hubbard $U$ correction (DFT+U)~\cite{Anisimov1997} which accounts for the on-site Coulomb repulsion of $5d$ electrons of Os. For monolayer OsCl$_3$, Fig.~\ref{fig:fig2}(d) reveals a possibility of a transition from QAH insulating phase with $\mathcal{C} \neq 0$ to Mott insulating  phase with $\mathcal{C}=0$ as $U$ is increased toward some critical value $U_c \simeq 0.35$ eV.

\begin{table}[t]
	\caption{The total energy $E_\mathrm{tot}$ per unit cell (in meV, relative to $E_\mathrm{tot}$ of FM$^x$ ground state), 
		as well as spin $\langle S \rangle$ and orbital $\langle O \rangle$ moments (in $\mu_B$), for several magnetic configurations of 
		Os atoms (first four of which are illustrated in Fig.~\ref{fig:fig4}) calculated by GGA+SOC method. Paramagnetic state has 
		$E_\mathrm{tot}=47.36$ meV and $\langle S \rangle = \langle O \rangle=0$.}\label{tab:magnet}
	\begin{tabular}{cccccccc}
		\hline
		                     & FM$^x$ & NAFM$^x$ & SAFM$^x$ & ZAFM$^x$ &FM$^z$ &FM$^y$ &NAFM$^y$\\
		\hline \hline
		E$_\mathrm{tot}$     &0.0     &39.60     &47.57     &16.63     &27.42   & 0.51  & 39.61 \\
		$\langle S \rangle$  &0.57    &0.13      &0.10      &0.41      &0.30    & 0.57  & 0.13  \\
		$\langle O \rangle$  &0.30    &0.24      &0.90      &0.33      & 0.12   & 0.30  & 0.24   \\
		\hline
	\end{tabular}
\end{table}

{\em Magnetic orderings of Os atoms.---}Surprisingly, despite requirement for large SOC to open sizable $E_g$, studies of candidate QAH insulators often na\"{i}vely assume that their magnetization is  perpendicular to the plane of 2D electron system~\cite{Liu2016}, thereby neglecting SOC-generated magnetocrystalline anisotropy (MCA). The MCA is fundamental property of any magnet which selects energetically favorable  magnetization  directions (such as easy-axis, easy-plane or easy-cone) and determines stability of that direction~\cite{Zhang2012c}. Moreover, MCA is crucial~\cite{Binder1976} to evade the Mermin-Wagner theorem according to which $T_\mathrm{C} \equiv 0$ for isotropic Heisenberg ferromagnet (with finite-range of interactions between its spins). 

Since in-plane magnetization or magnetic field cannot by itself induce quantized Hall conductance, some studies of potential QAH insulators have proposed~\cite{Zhang2012c} to apply external electric field to change MCA energy so that easy-axis tilts out of the plane. However, even {\em in-plane magnetization} can generate QAH effect in crystals with preserved inversion symmetry but broken out-of-plane mirror reflection symmetry (i.e., $z \rightarrow -z$)~\cite{Ren2016}, as satisfied by the lattice shown in Fig.~\ref{fig:fig1}. Another important issue is that ferromagnetic ordering, where all spins point in or out of the plane of OsCl$_3$, might not be the ground state of magnetic moments residing on the sites of the honeycomb lattice~\cite{Price2012}. 

Therefore, prior to analyzing electronic band structure of infinite and semi-infinite sheets, as well as nanowires, of OsCl$_3$, we first investigate total energy $E_\mathrm{tot}$ of four possible in-plane magnetic configurations shown in Fig.~\ref{fig:fig4}(a), as well as of the paramagnetic one and the ferromagnetic one but whose spins point out of the plane (FM$^z$). Table~\ref{tab:magnet} shows that ferromagnetic configuration FM$^x$, whose spins point along the $x$-axis, is the ground state (even at finite $U < U_c$). Additionally, Fig.~\ref{fig:fig4}(b) shows that energy $E_\mathrm{tot}$ increases if Os spins try to move away from the $x$-axis as the easy-axis. Thus, we conclude that monolayer OsCl$_3$ is an anisotropic 2D $XY$  ferromagnet~\cite{Ma1997}, or, more precisely, ``planar rotator''~\cite{Costa1996} ferromagnet as a special case of 2D $XY$ model where $S_z \equiv 0$. Its spin-spin interactions are described by an effective Hamiltonian $H_\parallel = - \sum_{\langle ij \rangle} J_\parallel (S_i^x S_j^x + S_i^y S_j^y) -  \sum_i D (S_i^x)^2$, where $(S_x,S_y)$ is the spin operator in quantum~\cite{Ma1997} or unit vector~\cite{Costa1996} in the classical version of the model; $\langle ij \rangle$ denotes the summation over nearest neighbors; and $D$ quantifies anisotropy ($D \rightarrow \infty$ suppresses fluctuations in the $y$-component of the spins, thereby leading to the Ising model). The 2D system described by $H_\parallel$ and sufficiently large $D/J_\parallel$ undergoes transition from ferromagnet to paramagnet at the Curie temperature which we estimate~\cite{Spirin2003} as $T_\mathrm{C} \lesssim 350$ K for $D/J \simeq 0.4$ according to Table~\ref{tab:magnet} and Fig.~\ref{fig:fig4}(b) [for $D/J \lesssim 10^{-7}$~\cite{Spirin2003}, one expects ferromagnet to transition to Berezinskii-Kosterlitz-Thouless (BKT) phase with topological order, and then to paramagnet at $T_\mathrm{BKT}$]. The interplane exchange coupling described by an additional Hamiltonian, $H_\perp = - \sum_{\langle ij \rangle} J_\perp (S_i^x S_j^x + S_i^y S_j^y)$, would further increase~\cite{Costa1996} $T_\mathrm{C}$ of OsCl$_3$ multilayers, whose spin Hamiltonian is given by $H_\parallel + H_\perp$ with $J_\perp/J_\parallel \simeq 0.1$ estimated from DFT calculations for OsCl$_3$ bilayer.

{\em Topology of bulk band structure.---}Figures~\ref{fig:fig2}(a) and ~\ref{fig:fig2}(c) show bulk band structure of monolayers OsCl$_3$ (assuming $U=0$) in the FM$^x$ and FM$^z$ configuration of Os spins, respectively, computed using DFT implemented in VASP  package~\cite{Kresse1993,Kresse1996a}. The electron core interactions are described by the projector augmented wave (PAW) method~\cite{Blochl1994,Kresse1999}, and we use Perdew-Burke-Ernzerhof parametrization of generalized gradient approximation (GGA) for the exchange-correlation functional.  The cutoff energy for the plane wave basis set is 500 eV for all calculations. The $k$-point mesh $10\times10\times1$ within Monkhorst-Pack scheme is used for the Brillouin zone integration in the self-consistency cycle. For the density of states (DOS) and $E_\mathrm{tot}$ in Table~\ref{tab:magnet} we use finer meshes of $25 \times 25 \times 1$ and $32 \times 32 \times 1$ points, respectively.  We fully optimize the atomic coordinates until Hellmann-Feynman forces on each ion are less than $0.01$  eV/{\AA}, which yields the lattice constant \mbox{$a=5.99$ \AA}.

When SOC is switched off, Fig.~\ref{fig:fig2}(a) shows that monolayer OsCl$_3$ exhibits nonzero exchange splitting between spin-up and spin-down bands, but remains metallic. Switching SOC on opens a  band gap $E_g \simeq 67$ meV around the Fermi energy $E_F$, as highlighted by the DOS in Fig.~\ref{fig:fig2}(b). Furthermore, this gap is topologically nontrivial as indicated by the nonzero Chern number, $\mathcal{C}=-1$ (note that we find $\mathcal{C}=3$ for VCl$_3$ and $\mathcal{C}=-1$ for RuCl$_3$, but $\mathcal{C}=0$ for FeCl$_3$, CoCl$_3$ and IrCl$_3$). Figure~\ref{fig:fig2}(c) shows that \mbox{$E_g \simeq 191$ meV} would be even larger for FM$^z$ configuration of spins, but this requires to apply very large out-of-plane electric field due to the large difference between $E_\mathrm{tot}$  of FM$^x$ and FM$^z$ configurations in Table~\ref{tab:magnet}. The Chern number $\mathcal{C}=1$ is obtained for FM$^z$ configuration of spins in Fig.~\ref{fig:fig2}(c), which also shows how $\mathcal{C}$ would change by moving $E_F$ (such as by applying gate voltage~\cite{Li2016}) into different gaps of the bulk band structure in Fig.~\ref{fig:fig2}(c).

We compute $\mathcal{C}$ from the Kubo formula~\cite{Thouless1982} using the $k$-space Hamiltonian obtained by Fourier transforming the real-space tight-binding Hamiltonian (TBH) in the basis of maximally localized Wannier functions~\cite{Marzari2012}. Since low-energy physics is mainly dominated by $5d$ orbitals of transition-metal atoms, we use all five $d$ orbitals centered on each Os atom, while $p$ orbitals on Cl atoms are neglected. The on-site potential and hopping through the fourth-nearest neighbor are calculated directly from GGA+SOC Hamiltonian generated by DFT calculations. Moreover,  the DOS in Fig.~\ref{fig:fig2}(b) shows that crystal field splits $d$ orbitals into $t_{2g}$ and $e_g$, where only three  $t_{2g}$ orbitals contribute to the DOS around $E_F$ (whereas the  contribution of $p$ orbitals of Cl is much smaller). Thus, TBH constructed from $t_{2g}$ orbitals alone centered on the sites of the honeycomb lattice would also give nonzero Chern number, even for  in-plane magnetization, because it does not commute with $z \rightarrow -z$ mirror reflection~\cite{Ren2016}.

We also use Wannier TBH to obtain the local DOS at the zigzag or armchair edge of a sheet of OsCl$_3$ which is infinite in the $x$-direction and semi-infinite in the $y$-direction. The local DOS,  obtained from the retarded Green function as $-\mathrm{Im} \, \hat{G}^r(E;k_x,j)/\pi$~\cite{Zhang2009}, is plotted in Fig.~\ref{fig:fig3}. It confirms the presence of one chiral edge state, whose E-$k_x$ dispersion penetrates through $E_g$ gap, in accord with $\mathcal{C} = \pm 1$ and {\em bulk-boundary correspondence}.

{\em Correlation-driven topological quantum phase transition.---}Although accurate value of $U$ is not known for OsCl$_3$, we expect moderate correlation effects due to spatially extended $5d$ orbitals of Os. Therefore, similarly to recent studies of electronic phases in other Os compounds~\cite{Wan2012} we vary $U$ from 0 to 1.5 eV in Fig.~\ref{fig:fig2}(d). This reveals that QAH insulating phase persists up to some critical value $U_c \simeq 0.35$ eV, at which $E_g \rightarrow 0$ due to electron correlations. Further increase of $U>U_c$ opens a gap of SO-coupled Mott insulating phase which is topologically trivial with $\mathcal{C}=0$. Thus, transition from QAH insulator to Mott insulator is an example of a continuous (due to gap closing at $U_c$) topological (due to change of topological number $\mathcal{C}$) quantum phase transition~\cite{Roy2016}.

{\em Spatial and transport properties chiral edge states.---} Using Wannier TBH we compute the real-space retarded Green function~\cite{Datta1995}, \mbox{$\hat{G}^r(E;i,j)=[E-\hat{H}-\hat{\Sigma}^r_L -\hat{\Sigma}^r_R]^{-1}$},  of the central region of OsCl$_3$ nanoribbons attached to the left (L) and right (R) semi-infinite leads of the same width $W$. The leads generate self-energies $\hat{\Sigma}^r_{L,R}$ determining escape rates of electrons from the central region into the macroscopic reservoirs kept at electrochemical potential $\mu_{L,R}$. The active region is either perfectly clean or host two vacancies (one on each edge) as defects obtained by cutting out all bonds adjacent to a chosen Os site. For example, in Dirac materials like graphene and Bi$_2$Se$_3$ vacancies drastically affect the local DOS, by introducing sharp peaks of quasilocalized character which can be detected by scanning tunneling microscopy~\cite{Wehling2014}, and represent infinitely strong scatterers which cannot be handled by any perturbative analysis~\cite{Ferreira2015}.

The subband structure of clean zigzag and armchair nanoribbons is shown in Figs.~\ref{fig:fig5}(a) and ~\ref{fig:fig5}(e), respectively. The transmission function~\cite{Datta1995}, $T(E)  =\mathrm{Tr}[\hat{\Gamma}_R \hat{G}^r \hat{\Gamma}_L (\hat{G}^r)^\dagger]$, at vanishing applied bias voltage \mbox{$\mu_L - \mu_R \rightarrow 0$} is plotted in Figs.~\ref{fig:fig5}(b) and ~\ref{fig:fig5}(f). It exhibits quantized steps in the clean case, which are destroyed outside of the bulk gap region in Figs.~\ref{fig:fig5}(a) and ~\ref{fig:fig5}(e) when vacancies are introduced. However, $T(E)$ in Fig.~\ref{fig:fig5}(f) has a dip even within the topologically protected bulk gap due to wide spatial extension of edge states in the armchair case, so that hybridization of edge states from opposite edges opens a minigap at the crossing point in Fig.~\ref{fig:fig5}(e). The local DOS, $-\mathrm{Im} \, \hat{G}^r(E;i,i)/\pi$, shown for zigzag in Fig.~\ref{fig:fig5}(c) and armchair nanoribbons in Fig.~\ref{fig:fig5}(g) confirms that edge states in the former case are narrower than in the latter case, as observed previously~\cite{Chang2014,Prada2013} for edge states of QSH insulators emerging from different types of graphene nanoribbons. 

The zigzag edge changes the Hamiltonian near the boundary which then introduces a kink in the dispersion of edge state in Figs.~\ref{fig:fig3}(c) and ~\ref{fig:fig5}(a), as also found for Haldane model defined on zigzag nanoribbon~\cite{Kane2013}. For some values of energy $E$ within the bulk band gap, such dispersion intersect with $E$ at $N_R=2$ points with positive velocity and $N_L=1$ points with negative velocity. This leads to $T(E)=3$ for $E$ within a portion of the bulk band gap in Fig.~\ref{fig:fig5}(b). Nevertheless, only the difference $N_R - N_L=\mathcal{C}$ is topologically protected according to {\em bulk-boundary correspondence}~\cite{Kane2013}, leading to quantized $T(E)=1$ across the whole bulk band gap even in the presence of vacancies in Fig.~\ref{fig:fig5}(b).

{\em Conclusions.---}Using DFT and DFT+U, as well as quantum transport and MCA energy, calculations, we predict that monolayer of the layered material OsCl$_3$ offers a playground to examine electronic phases governed by the {\em interplay between SOC, magnetic orderings with large anisotropy due to strong SOC and correlations of $5d$ electrons}. Upon increasing the on-site Coulomb repulsion, monolayer OsCl$_3$ undergoes quantum phase transition from QAH insulator to correlated metal, and finally to topologically trivial SO-coupled Mott insulator. Interestingly, large gap $E_g=67$ meV with nonzero Chern number $\mathcal{C}=-1$ is observed even though its spontaneous magnetization is along the in-plane easy-axis, which is possible when the mirror reflection symmetry is broken~\cite{Ren2016}.  We note that Ref.~\cite{Witczak-Krempa2014} has proposed a heuristic phase diagram in which QAH insulating phase borders~\cite{Song2016}  a trivial Mott insulator. Thus, our Fig.~\ref{fig:fig2}(d) can be viewed as a prescription for realizing such phase diagrams by using a realistic materials whose 2D nature makes possible  to manipulate its charge density and Fermi energy by the gate electrode (as demonstrated very recently in correlated monolayer TMDs~\cite{Li2016}). The correlations of $5d$ electrons characterized by spatially extended wave functions can be captured more accurately with DFT+dynamical mean field theory~\cite{Held2007}, so we relegate re-examination  of the phase diagram in Fig.~\ref{fig:fig2}(d) via this methodology for future studies.

\begin{acknowledgments}
	 We thank Jiadong Zang for valuable insights. This work was supported by NSF Grant No. ECCS 1509094. The supercomputing time was provided by XSEDE, which is supported by NSF Grant No. ACI-1053575.
\end{acknowledgments}



\end{document}